\documentclass[prd,aps,tightenlines,amsmath,amssymb,12pt,nofootinbib,floatfix,showkeys,showpacs]{revtex4}

\usepackage{graphicx}
\usepackage{dcolumn}
\usepackage{bm}

\usepackage{natbib}
\bibpunct{(}{)}{;}{a}{}{,} 

\def\ve#1{{\mbox{\boldmath$#1$}}}
\begin{document}

\title{Explicit exact expression for the Thomas precession}

\author{Sergei A. \surname{Klioner}}

\affiliation{Lohrmann Observatory, Dresden Technical University,
Mommsenstr.~13, 01062 Dresden, Germany\\
}

\begin{abstract}
\begin{center}
\bf
GAIA-CA-TN-LO-SK-004-2

issue 2, 14 March 2008
\end{center}

This work gives an explicit exact expression for the Thomas precession
arising in the framework of Special Theory of Relativity as the spatial
rotation resulting from two subsequence Lorentz boosts. The final
result for the orthogonal matrix of Thomas precession is given by Eqs.
(\ref{P-exact})--(\ref{D}). A trivial calculation leads to the compact
formula (\ref{P-angle}) for the angle of rotation due to Thomas
precession.

In the framework of Gaia the special-relativistic Thomas precession is
an important step in the derivation of an aberrational formula with the
Mansouri-Sexl parameters. The latter formula will be used to test the
Local Lorentz Invariance with Gaia data as will be explained elsewhere.
\end{abstract}


\maketitle

\section{Introduction}

The Thomas precession naturally arises in Special Theory of Relativity
as the additional rotation to be added to a Lorentz boost to represent
the result of two subsequence Lorentz boosts. Although the derivation
of the Thomas precession from the Lorentz transformations can be found
in many textbooks (see, e.g., \citet{Jackson1975} or
\citet{Moeller1972}), it is normally done in the form of expansion in
powers on $1/c$. \citet{Salingaros1986} gives the exact expression for
Thomas precession, but in a form that is not readily useful for further
calculations. \citet{SexlUrbantke2001} have also given the exact
expression, but have not simplified it algebraically, leaving the reader
with a rather lengthy calculations. The purpose of this short note is
to derive the exact and fully simplified expression for the Thomas
precession directly using two subsequence Lorentz transformations and
representing them as a Lorentz transformation plus a spatial rotation.
All calculations have been performed explicitly and in normal vector
notations. The resulting formula for Thomas precession (Eqs.
(\ref{P-exact})--(\ref{D}) below) is rather compact and is valid
exactly. Using this expression for the Thomas precession a trivial
calculation leads to Eq. (\ref{P-angle}) for the angle of rotation due
to Thomas precession.

The exact formula for the Thomas precession is interesting by itself,
but can also be considered as a step in the discussion of the Thomas
precession in the framework of Mansouri-Sexl test theory
\citep{MansouriSexl1977}. That latter discussion is important to
interpret the results of various modern project performing
high-accuracy directional measurements (e.g., Gravity Probe B, Gaia or
SIM) in terms of the Local Lorentz Invariance \citep[see, e.g.,][for
the case of Gaia and SIM]{Klioner2007}.

The notations of this paper are usual: $c$ is the velocity of light in
vacuum, lowercase Latin indices take values $1$, $2$, and $3$ and refer
to spatial components of corresponding quantities, index $0$ is used
for time components, Greek indices take values $0$, $1$, $2$ and $3$
and refer to all space-time components of corresponding quantities,
repeated indices (both Latin and Greek ones) imply Einstein summation
rule irrespective of their positions (e.g., $a_i\, b_i = a_1 b_1 + a_2
b_2 + a_3 b_3)$, the spatial components of a quantity considered as a
3-vector are set in boldface ($\ve{a} = a^i$), the absolute value
(Euclidean norm) of a 3-vector $\ve{a}$ is denoted as $a$ or $|\ve{a}|$
and is defined by  $a=|\ve{a}| = \left(a^1\,a^1 + a^2\,a^2 +
a^3\,a^3\right)^{1/2}$, the scalar product of any two 3-vectors
$\ve{a}$ and $\ve{b}$ with respect to the Euclidean metric
$\delta_{ij}$ is denoted as $\ve{a} \cdot \ve{b}$ and defined as
$\ve{a} \cdot \ve{b} = \delta_{ij}a^i\,b^j = a^i\,b^i$, the Kronecker
symbol (unit matrix) is denoted as $\delta^{ij}$, parentheses
surrounding a group of indices denote symmetrization (e.g.,
$A^{(ij)}={1\over 2}\left(A^{ij}+A^{ji}\right)$), brackets surrounding
two indices denote antisymmetrization (e.g., $A^{[ij]}={1\over
2}\left(A^{ij}-A^{ji}\right)$).

\section{Two subsequent Lorentz transformations and the Thomas precession}

Let us consider three inertial reference systems: $(X^0=c\,T,X^i)$,
$(x^0=c\,t,x^i)$ and $(\hat{x}^{0}=c\,\hat{t},\hat{x}^i)$. The
velocity of $x^\alpha$ with respect to $X^\alpha$ is $V^i$. The
coordinates $X^\alpha$ and $x^\alpha$ are related by a Lorentz
transformation of the form
\begin{equation}
\label{x-X}
x^\alpha=\Lambda^\alpha_\beta\,X^\beta
\end{equation}
\noindent
where
\begin{eqnarray}
\label{eq:l1}
\Lambda^0_0 &=& \Gamma ,
\\
\label{eq:l2}
\Lambda^0_a &=& -\Gamma\,K^a ,
\\
\label{eq:l3}
\Lambda^i_0 &=& -\Gamma\,K^i ,
\\
\label{eq:l4}
\Lambda^i_a &=& \delta^{ia} + \frac{\Gamma^2}{1+\Gamma}\,K^i\,K^a ,
\\
\label{eq:l5}
\Gamma &=& \left(1-\ve{K}\cdot\ve{K}\right)^{-1/2} ,
\\
\label{eq:l6}
\ve{K} &=& \frac{1}{c}\,\ve{V} .
\end{eqnarray}
\noindent
The inverse transformation reads
\begin{equation}
\label{X-x}
X^\alpha=\widetilde\Lambda^\alpha_\beta\,x^\beta,
\end{equation}
\noindent
where $\widetilde\Lambda^\alpha_\beta$ is equal to
$\Lambda^\alpha_\beta$ with $-\ve{K}$ substituted for $\ve{K}$
($\Gamma$ remains the same after this substitution). The velocity of
reference system $\hat{x}^\alpha$ with respect to $x^\alpha$ is
$v^i$, and one has
\begin{equation}
\label{x'-x}
\hat{x}^\alpha=\lambda^\alpha_\beta\,x^{\beta},
\end{equation}
\noindent
where $\lambda^\alpha_\beta$ has the same form as $\Lambda^\alpha_\beta$ with
\begin{eqnarray}
\label{gamma}
\gamma &=& \left(1-\ve{k}\cdot\ve{k}\right)^{-1/2} ,
\\
\label{k}
\ve{k} &=& \frac{1}{c}\,\ve{v}
\end{eqnarray}
\noindent
substituted for $\Gamma$ and $\ve{K}$, respectively.
Now, the velocity of $\hat{x}^\alpha$ relative to $X^\alpha$ is $\hat{\ve{V}}$.
Using standard considerations one gets the relation between
the three velocities:

\begin{equation}
\label{Kprime-k-K}
\hat{\ve{K}}={1\over 1+p}\,\left[\ {1\over \Gamma}\ \ve{k}+\left(1+{\Gamma\over 1+\Gamma}\,p\right)\,\ve{K}\right],
\end{equation}
\noindent
where
\begin{eqnarray}
\label{K-prime}
\hat{\ve{K}} &=& \frac{1}{c}\,\hat{\ve{V}},
\\
\label{p}
p&=&\ve{k}\cdot\ve{K}.
\end{eqnarray}
\noindent
Combining (\ref{x-X}) and (\ref{x'-x}) one has the relation between $\hat{x}^\alpha$ and $X^\alpha$:
\begin{eqnarray}
\hat{x}^\alpha&=&\Sigma^{\alpha}_{\ \beta}\,X^\beta,
\\
\Sigma^{\alpha}_{\ \beta}&=&\lambda^{\alpha}_{\ \rho}\,\Lambda^{\rho}_{\ \beta}.
\end{eqnarray}
\noindent
Now, let us define matrix
$\hat{\Lambda}^\alpha_\beta$ with the same structure as $\Lambda^\alpha_\beta$ but with
\begin{eqnarray}
\label{Gamma-prime}
\hat{\Gamma} &=& \left(1-\hat{\ve{K}}\cdot\hat{\ve{K}}\right)^{-1/2}
\end{eqnarray}
\noindent
substituted for $\Gamma$ and $\hat{\ve{K}}$ for $\ve{K}$. According to
(\ref{Gamma-prime}), (\ref{gamma}), (\ref{eq:l5}) and
(\ref{Kprime-k-K}) one gets
\begin{equation}
\hat{\Gamma}=\gamma\,\Gamma\,(1+p).
\end{equation}
\noindent
Straightforward calculations show that
\begin{eqnarray}
\Sigma^{0}_{\ \beta}&=&\hat{\Lambda}^0_\beta,
\\
\Sigma^{a}_{\ \beta}&=&P^{ab}\,\hat{\Lambda}^b_\beta,
\end{eqnarray}
\noindent
where $P^{ab}$ is the orthogonal matrix describing the Thomas precession
\begin{eqnarray}
\label{P-exact}
P^{ab}&=&\delta^{ab}+{\cal A}\,K^a\,K^b+{\cal B}\,k^a\,K^b+{\cal C}\,K^a\,k^b+{\cal D}\,k^a\,k^b\,,
\\
\label{A}
{\cal A}&=&{(1-\gamma)\,\Gamma^2\over(1+\Gamma)\,(1+\hat{\Gamma})}\,,
\\
\label{B}
{\cal B}&=&{\gamma\,\Gamma\over 1+\hat{\Gamma}}\,\left(1+2\,{\hat{\Gamma}-\gamma\,\Gamma\over(1+\gamma)\,(1+\Gamma)}\right)\,,
\\
\label{C}
{\cal C}&=&-{\gamma\,\Gamma\over 1+\hat{\Gamma}}\,,
\\
\label{D}
{\cal D}&=&{\gamma^2\,(1-\Gamma)\over(1+\gamma)\,(1+\hat{\Gamma})}\,.
\end{eqnarray}
\noindent
Matrix $P^{ab}$ is orthogonal and satisfies the relation
$P^{ac}\,P^{bc}=\delta^{ab}$. If $\ve{k}$ is parallel to $\ve{K}$ (that
is, for $\ve{k}=\alpha\,\ve{K}$ with any $\alpha$), it is easy to check
from (\ref{P-exact})--(\ref{D}) that the Thomas precession vanishes and
$P^{ab}=\delta^{ab}$. \citet{SexlUrbantke2001} have derived this
result, but have not given it in explicit and fully simplified form.

The angle of rotation $\alpha$ due to Thomas precession can be
directly computed from the trace of matrix $P^{ab}$ using the standard
formula $1+2\cos\alpha=P^{aa}$. Trivial calculation leads immediately to
\begin{equation}
\label{P-angle}
1+\cos\alpha={{\left(1+\gamma+\Gamma+\hat{\Gamma}\right)}^2\over(1+\gamma)\,(1+\Gamma)
\,(1+\hat{\Gamma})}.
\end{equation}
\noindent
The equivalent results have been derived after lengthy calculations by
\citet[][Eq. (124)]{Macfarlane1962} and \citet[][the last equation of
the paper]{Urbantke1990} and discussed also by \citet[][Eq.
(2.10.7)]{SexlUrbantke2001}.

\section{Important limits for the Thomas precession}

From this matrix one can easily restore all standard results concerning
the Thomas precession. Expanding $P^{ab}$ in terms of $k=|\ve{k}|$ one
gets

\begin{equation}
\label{exp1}
P^{ab}=\delta^{ab}+{2\,\Gamma\over 1+\Gamma}\,k^{[a}\,K^{b]}+{\cal O}(k^2),
\end{equation}
\noindent
where $A^{[i}B^{j]}={1\over 2}\left(A^iB^j-A^jB^i\right)$ is the
antisymmetric part of $A^i\,B^j$ for any two vectors $\ve{A}$ and
$\ve{B}$. Defining $\delta\ve{K}=\hat{\ve{K}}-\ve{K}$ and using
(\ref{Kprime-k-K}), Eq. (\ref{exp1}) can be re-written as
\begin{equation}
\label{exp2}
P^{ab}=\delta^{ab}+{2\,\Gamma^2\over 1+\Gamma}\,\delta K^{[a}\,K^{b]}+{\cal O}(|\delta\ve{K}|^2).
\end{equation}
\noindent
This latter form can be found, e.g., in \citet{Jackson1975} and \citet{Moeller1972}.
Finally, expanding (\ref{P-exact})--(\ref{D}) in powers of $1/c$ one gets
\begin{eqnarray}
\label{exp3}
P^{ab}&=&\delta^{ab}+k^{[a}\,K^{b]}
+{1\over 4}\,k^{[a}\,K^{b]}\,\left(k^2+K^2-p\right)
\nonumber\\
&&
-{1\over 8}\,\left(k^2\,K^a\,K^b+K^2\,k^a\,k^b-2\,p\,k^{(a}\,K^{b)}\right)
\nonumber\\
&&
+{\cal O}(c^{-6}).
\end{eqnarray}
\noindent
This expansion can be conveniently used for modelling of high-accuracy
directional data. Let us note that the symmetric part of the terms of
order ${\cal O}(c^{-4})$ immediately follows from the antisymmetric
terms of order ${\cal O}(c^{-2})$. Indeed, considering a general
representation
$P^{ab}=\delta^{ab}+\sum_{k=1}^\infty\epsilon^k\,f^{ab}_k$ with any
formal parameter, the condition of orthogonality
$P^{ac}\,P^{bc}=\delta^{ab}$ allows one to determine the symmetric term
$f^{(ab)}_k$ at any order of $\epsilon$. In particular one has
$f^{(ab)}_{1}=0$, $f^{(ab)}_{2}=-{1\over
2}\,f^{[ac]}_{1}\,f^{[bc]}_{1}$, $f^{(ab)}_{3}=-{1\over
2}\,\left(f^{[ac]}_{1}\,f^{[bc]}_{2}+f^{[bc]}_{1}\,f^{[ac]}_{2}\right)$.
With $\epsilon\,f^{[ab]}_{1}=k^{[a}\,K^{b]}$ one
immediately restores the symmetric part of the terms of order ${\cal
O}(c^{-4})$ given in the second line of (\ref{exp3}):
$\epsilon^2\,f^{(ab)}_{2}=-{1\over 8}\,\left(k^2\,K^a\,K^b+K^2\,k^a\,k^b-2\,p\,k^{(a}\,K^{b)}\right)$.

\acknowledgments

This work was partially supported by the BMWi grant 50\,QG\,0601
awarded by the Deutsche Zentrum f\"ur Luft- und Raumfahrt e.V. (DLR).

\end{document}